\documentclass[
pra,
nofootinbib,superscriptaddress,showkeys,showpacs,
preprintnumbers, 
amsmath,amssymb]{revtex4-2}

\usepackage{graphicx,color}
\usepackage{bm}
\usepackage{graphicx}
\usepackage{dcolumn}
\usepackage{bm}
\usepackage{color}
\usepackage{soul}
\usepackage{tikz}
\usepackage[bookmarksnumbered=true]{hyperref} 
\hypersetup{colorlinks = true,linkcolor = blue,anchorcolor = blue,citecolor = blue, filecolor = blue,urlcolor = blue}
\usetikzlibrary{shapes,arrows,positioning,fit,backgrounds,calc}
\usetikzlibrary{shapes.geometric, arrows}

\begin{document}
\preprint{APS/123-QED}

\title{Quantum entanglement between neutrino eigenstates in the presence of the subsequent phase shift of the neutrino oscillations}

\author{Hoda Abdolalizadeh} 
\email{abdolalizadehhoda@gmail.com}

\affiliation{Theoretical and Computational Physics Research Laboratory, Istanbul University, 34134 İstanbul, Türkiye}

\affiliation{Institute of Graduate Studies in Science, İstanbul University, 34134, İstanbul, Türkiye}

\author{Ekrem Aydiner}
\email{ekrem.aydiner@istanbul.edu.tr \\ https://orcid.org/0000-0002-0385-9916}

\affiliation{Theoretical and Computational Physics Research Laboratory, Istanbul University, 34134 İstanbul, Türkiye}

\affiliation{Department of Physics, Faculty of Science, İstanbul University, 34134 İstanbul, Türkiye}

\date{\today}

\begin{abstract}

In this Letter, using von Neumann entropy we examine the entanglement entropy for the neutrino oscillations in the presence of the subsequent phase shift. We numerically show that the entanglement entropy for the subsequent periods of the two-flavor neutrino oscillations increases asymmetrically with time depending on the space-time deformation. We also explored the obtained results for the three-flavor neutrino oscillations to show that this result is also valid for the three-flavor neutrino oscillations. These results, obtained for the first time in this Letter, are quite different from the computing for the standard neutrino oscillation theory. We concluded that these interesting results play an important role in the cosmology.

\end{abstract}

\maketitle


\section{Introduction}

Recently, quantum entanglement and the phase shift in neutrino oscillations have attracted much attention in neutrino physics. In this Letter, we will discuss two issues together. Before delving into the subject, we would like to briefly present the work done in both fields.

Entanglement entropy as Measuring the degree of entanglement is a fundamental concept in quantum mechanics. In the case of neutrino particles, the entanglement entropy indicates the degree of entanglement between the eigenstates of the neutrino mass. Considering the importance of various aspects of entanglement entropy in neutrino oscillations, many theoretical studies have addressed this issue.
It can be seen in the literature that Blasano et al have done a series of studies on quantum entanglement in neutrino oscillations. For instance, Blasone \textit{et al} quantified in detail the amount and the distribution of entanglement in the physically relevant cases of flavor mixing in quark and neutrino systems \cite{Blasone2008a}. In 2009, they showed that mode entanglement can be expressed in terms of flavor transition probabilities \cite{Blasone2009}. In 2010, they studied single-particle entanglement arising in two-flavor neutrino mixing, first in the context of quantum mechanics, and then in the case of quantum field theory \cite{Blasone2010}. In 2013, they described neutrino oscillations in terms of (dynamical) entanglement of neutrino flavor modes \cite{Blasone2013}. In 2014, they extended their studies to the framework of quantum field theory and used concurrence as a suitable measure of entanglement \cite{Blasone2014}, in another work, they analyzed the entanglement in the states of flavor neutrinos and antineutrinos within the framework of quantum information theory and, revealed its correlation with experimentally measurable quantities such as the variances of lepton numbers and charges \cite{Blasone2014b}. In 2015, they employed two measures: the concurrence and the logarithmic negativity, to quantify the entanglement, and they analyzed the behavior of multipartite entanglement within the (three-qubit) Hilbert space of flavor neutrino eigenstates \cite{Blasone2015}. Similarly, Kayser \textit{et al} examined the role of entanglement in theoretical models of neutrino oscillation, and they found that a theoretical approach considering the entanglement between neutrinos and their interaction partners still produces the standard result for the neutrino oscillation wavelength \cite{Kayser2010}. Martin \textit{et al} \cite{Martin2022} investigated measurements of quantum information, including entanglement entropy and purity in quantum neutrino oscillations at macroscopic scales in extreme astrophysical environments, such as the early universe, core-collapse supernovae, and merging neutron stars. Siwach \textit{et al} examined entanglement in neutrino systems by quantifying entropies and polarization vector components in the context of three-flavor oscillations \cite{Siwach}. Amol \textit{et al} presented a significant connection between the entanglement entropy of individual neutrinos and the occurrence of spectral splits in their energy spectra due to collective neutrino oscillations \cite{Amol}. Mallick \textit{et al} explored the entanglement entropy between spins and position space during neutrino propagation \cite{Mallick}. Cervia \textit{et al} measured the entanglement entropy and the Bloch vector of the reduced density matrix, which are used to assess the interactions between constituent neutrinos in the many-body system \cite{Cervia_2019}.

On the other hand, investigating the phase shift of neutrino oscillations, such as the entanglement entropy, is a fascinating topic proposed by Ahluwalia and Burgard \cite{Ahluwalia1996, Ahluwalia1998}. They theoretically showed that the effect of gravity appears in the form of phase shift in neutrino oscillations. After these seminal works, this concept has received attention in various studies within the literature. For instance, Grossman \textit{et al} argued that the gravitational oscillation phase shift in the neutrino oscillation might have a significant effect on supernova explosion \cite{Grossman1997}. Capozziello \textit{et al} discussed the gravitational phase shift of neutrino oscillation in the framework of $f(R)$-gravity \cite{CAPOZZIELLO_2010}. Additionally, Capozziello and  Lambiase analyzed the flavor oscillations of neutrinos in the framework of Brans-Dicke theory of gravity \cite{CAPOZZIELLO_1999}. Ren and Zhang presented general equations for oscillation phases under the influence of gravitational rotation and electric charge in Kerr-Newman spacetime \cite{Ren2010}. Koranga showed that there would be a phase shift in neutrino oscillation due to quantum gravity effects at the Planck scale \cite{Koranga2012}. In a similar study, Swami computed the phase shift in neutrino oscillations due to the rotation of gravitational sources for both non-lensed and lensed neutrinos Ref.\,\cite{Swami2022}. More recently, Aydiner theoretically proposed that phase shift between eigenstates may gradually increase in the neutrino oscillation \cite{Aydiner}. This means that phase shift in Aydiner model is not constant, unlike, the phase shift increases each subsequent oscillation period depending on the deformation of the space-time.

In this Letter, we will investigate the entanglement entropy between neutrino eigenstates in the presence of the subsequent phase shift based on Ref.\cite{Aydiner}. We will show that quantum entanglement between eigenstates also increases depending on the deformation of the space-time.

This work is organized as follows: In Section II, we first analyze the neutrino flavor oscillation model and then delve into the concept of entanglement entropy in neutrino oscillations. Subsequently, we calculate the entropy of neutrino oscillations between two flavors using Von Neumann's method. Section III, we present the impact of varying degrees of space-time deformation on the entanglement entropy and phase shift of neutrino oscillations. Finally, in the last section, a discussion and conclusion are given.

\section{Neutrino Oscillation and Quantum Entanglement} \label{Section-II}

\subsection{Neutrino Oscillation}

Neutrino oscillation is the probability of transition from a flavor to the other one during the neutrino propagation. The discrepancy between mass and flavor eigenstates and also the non-zero and non-degenerate neutrino mass give rise to the neutrino oscillation. Neutrino oscillation can be characterized by three mixing angles and a phase. The flavor eigenstates of the neutrino denoted by $|\nu_{\alpha} \rangle$ can be represented as linear superposition of the mass eigenstates $|\nu_{j} \rangle $
\begin{equation} \label{neutrino-state}
	|\nu_{\alpha} \rangle = \sum_{j} U^{*}_{\alpha j} |\nu_{ j} \rangle, \quad \alpha=e,\mu,\tau, \quad j=1,2,3
\end{equation}
where $U$ is a unitary and non-diagonal mixing matrix that specifies the composition of each neutrino flavor state. It can be given in the general form 
\begin{equation} \label{U-matix}
	U=\begin{pmatrix}
	c_{12} c_{13} & s_{12} c_{13} & s_{13} e^{-i\delta_{CP}}\\
        -s_{12} c_{23} - c_{12} s_{23} s_{13} e^{i\delta{CP}} & c_{12} c_{23} - s_{12} s_{23} s_{13} e^{i\delta{CP}} & s_{23} c_{13} \\
          s_{12} s_{23} - c_{12} c_{23} s_{13} e^{i\delta{CP}} & -c_{12} s_{23}- s_{12} c_{23} s_{13} e^{i\delta{CP}} & c_{23} c_{13} \\
	\end{pmatrix}
\end{equation}
where $c_{ij}=\cos \theta_{ij}$, $s_{ij}=\sin \theta_{ij}$ and $\delta$ is the Dirac CP violating phase. This matrix is called the Pontecorvo–Maki–Nakagawa–Sakata (PMNS) matrix \cite{Gribov1969,Maki1962}. The PMNS matrix represents the transformation between flavor and mass eigenstates, providing insights into how the probabilities of different flavor outcomes are entangled with the underlying mass states. For three flavors Eq.(\ref{neutrino-state}) can be written as
\begin{equation}\label{3flavors-eigenstate}
    \begin{pmatrix}
        \nu_e\\
        \nu_\mu\\
        \nu_\tau\\
    \end{pmatrix}
    =
   \begin{pmatrix}
       U_{e1} & U_{e2} & U_{e3} \\
       U_{\mu1} & U_{\mu2} & U_{\mu3}\\
       U_{\tau1} & U_{\tau2} & U_{\tau3}\\
   \end{pmatrix}
    \begin{pmatrix}
        \nu_1\\
        \nu_2\\
        \nu_3\\
    \end{pmatrix}
\end{equation}
However, Eq.(\ref{3flavors-eigenstate}) for two flavors can be given in the simple form: 
\begin{equation}\label{2flavors-eigenstate}
    \begin{pmatrix}
        \nu_e\\
        \nu_\mu\\
    \end{pmatrix}
    =
    \begin{pmatrix}
        \cos\theta & \sin\theta\\
        -\sin\theta & \cos\theta
    \end{pmatrix}
    \begin{pmatrix}
        \nu_1\\
        \nu_2\\
    \end{pmatrix}
\end{equation}
where $\nu_e$ and $\nu_\mu$ can be given in the eigenstate forms
 \begin{equation}
     {|\nu_e(t=0)\rangle} =\cos{\theta}{|\nu_1\rangle} +\sin{\theta}{|\nu_2\rangle} , \qquad
     {|\nu_\mu(t=0)\rangle} =-\sin{\theta} {|\nu_1\rangle} +\cos{\theta}{|\nu_2\rangle}.
\end{equation}

These equations describe the initial states of electron neutrinos and muon neutrinos as linear combinations of the flavor eigenstates $|\nu_1\rangle$ and $|\nu_2\rangle$. The weak eigenstates are rotated by an angle $\theta$ with respect to the mass eigenstates ${|\nu_1\rangle}$ and ${|\nu_2\rangle}$ to allow mixing between $\nu_\mu$ and $\nu_e$. The time-evolved state of flavor eigenstate $\nu_\alpha$, with $\alpha=e$, $\mu$ is a superposition of the mass eigenstates $\nu_k$, with $k=1$ and 2,
\begin{equation}
     {|{\nu}_{\alpha}(x ,t)\rangle} = \sum_{k} U_{\alpha k}\exp[-i\phi_k] {|{\nu}_k\rangle}
     \end{equation}
where the phase factor is 
\begin{equation}\label{phase factor}
    \phi_k = E_k t -\vec{P}_k . \vec{x}
\end{equation}
where $E_k$ and $\vec{P}$ are the energy and momentum of the mass eigenstates ${|{\nu}_k\rangle}$. The oscillation probability that the neutrino produced
as ${|\nu_e\rangle}$ is detected as ${|\nu_\mu\rangle}$ is 
\begin{equation}
    P_({\nu_e}\rightarrow{\nu_\mu}) = {\mid|\nu_e\rangle\langle \nu_\mu (x,t)|\mid}^2 = \sin^2{2\theta} \sin^2{\frac{\phi_{21}}{2}}
\end{equation}

If the neutrino produced at a space-time point $A(t_A ,\vec x_A)$ and detected at $B(t_B ,\vec x_B)$, the expression for the phase \cite{Follin}:
\begin{equation}\label{phase}
    \phi_{21} = (\frac{m_2 ^2}{2E_2}- \frac{ m_1 ^2}{2E_1}) \mid \vec x_B -\vec x_A \mid 
\end{equation}
$ \mid \vec x_B -\vec x_A \mid = L$, L is the length of the source from the detector also known as the \textit{baseline}; Usually this distance is much larger than the sizes of the production and detections regions \cite{Akhmedov2011}.
\begin{equation}\label{probability}
    P_({\nu_e}\rightarrow{\nu_\mu}) = \sin^2{2\theta} \sin^2{\frac{\Delta m^2 L}{4 E_0}}
\end{equation}
The probability of a neutrino oscillating from one flavor to another is given by a sinusoidal function, and it depends on parameters like the baseline distance traveled by the neutrinos, the energy of the neutrinos, and the mass-squared differences between the mass eigenstates. It should be noted that the functional dependence on $L/E$ is called a \textit{spectral dependence} \cite{Upadhyay,Kajita,Kimura}.

However, Aydiner proposed that the Dirac equation could be written in fractional form in the case of the neutrinos being coupled to deformed space-time \cite{Aydiner}:
 \begin{equation}\label{deformed-Dirac}
     iD_{{t}}^{\eta}\Psi_j(t,x)=- \left[ |P|  + \frac{(m_j)^2 c^3}{2 |P|}  \right] \frac{\delta.P}{2 |P|} \Psi_j(t,x)
 \end{equation}
where $D_{{t}}^{\eta}$ denotes the Caputo fractional derivative operator, $\eta$ denotes coupling strength or deformation of the space-time, which is given by $0 <\eta < 1$ and $\Psi(t,x)$ is the wave function of the neutrino. This equation leads to a fractional transition probability between quantum states $|\nu_\alpha\rangle \rightarrow\ |\nu_\beta\rangle$. This probability is given by
\begin{equation}\label{defomed-probability}
P_{|\nu_\alpha\rangle \rightarrow\ |\nu_\beta\rangle} =  \sum_{j,k} U_{\alpha j} U_{\beta j^*} U_{\alpha k^*} U_{\beta k} \exp \left[ -i \left(\frac{(m_j ^2 - m_k ^2)c^3}{ 2h |P| } \right)^\eta t^\eta \right ]
\end{equation}
which indicates anomalous cyclic behavior in the neutrino oscillation \cite{Aydiner}. This model provides that the phase shift will not be constant as in previous models, but the phase shift increases with each subsequent oscillation.  Additionally, it should be noted that in the Pontecorvo’s oscillation scheme, oscillation between mass and flavour eigenstates of the neutrino is the periodic. This Markovian behavior of the oscillation can be represented by Eq.(\ref{probability}). However, Aydiner \cite{Aydiner} suggested that the gravitational waves, the mass distribution in the universe or decoherence and deformation of the Hilbert space at the quantum level can cause the space-time deformations at both the local and cosmological scales. In this case, if the neutrino mass or flavor eigenvector is coupled to deformed space-time, time-dependent equation of the neutrino oscillation can be given in fractional form as in Eq.(\ref{deformed-Dirac}).

\subsection{Quantum Entanglement Between Neutrino Eigen-States} 

It is known that von Neumann entropy serves as a fundamental tool for quantifying entanglement between subsystems in bipartite quantum systems. When dealing with mixed states, where subsystems are described by density matrices, Von Neumann entropy, denoted as $S(\rho)$, offers a precise measure of entanglement by characterizing the degree of correlation between the subsystems. The density matrix is a mathematical representation that describes the quantum state of a system, taking into account the probabilities of different possible states, and the reduced density matrix is obtained by tracing out (or integrating over) the degrees of freedom of one of the subsystems from the full density matrix of the entire system \cite{Amico}. The Von Neumann entropy is defined as follows:

\begin{equation}\label{entropy1}
    S(\rho) = -\text{Tr}(\rho \ln \rho)
\end{equation}
where $\rho$ is the density matrix of neutrino flavor states (density matrices are normalized so that $\text{Tr} \rho=1$):
\begin{equation}\label{density matrix}
    \rho = {|{\nu}_{\alpha}(t)\rangle} {\langle {\nu}_{\alpha}(t)|}
\end{equation}

The density matrix is a mathematical object that contains all of the information about the state of the system. We limit ourselves to two flavors $\nu_{\alpha}$ and $\nu_{\beta}$, we can take them as the upper and lower components in the neutrino flavor iso-spin SU(2) algebra \cite{Balantekin_2022}, So by assuming the neutrino occupation number associated with a given flavor (mode) as a reference quantum number, one can establish the following correspondences with two-qubit states:

\begin{equation}\label{flavor-state}
    {|{\nu}_{\alpha}\rangle}\equiv|{1 \,\,0}\rangle=\begin{pmatrix}
        0\\
        1\\
        0\\
        0
    \end{pmatrix}, \quad  {|{\nu}_{\beta}\rangle}\equiv|{0 \,\,1}\rangle = \begin{pmatrix}
        0\\
        0\\
        1\\
        0
    \end{pmatrix}
\end{equation}

States $|0\rangle$ and $|1\rangle$ correspond, respectively, to the absence and the presence of a neutrino in mode $\alpha$ or $\beta$. Entanglement is thus established among flavor modes, in a single-particle setting.
The free evolution of the neutrino state ${|{\nu}_{\alpha}\rangle}$ can be written in the form:
\begin{equation}\label{evolution of flavor1}
    {|{\nu}_{\alpha}(t)\rangle} =  \sum_{j} U_{\alpha j} \exp^{\frac{-iE_j t}{h}}(\sum_{\alpha}U^{-1}_{\alpha j}{|{\nu}_{\alpha}\rangle})
\end{equation}
explicit expression of the Eq.(\ref{evolution of flavor1}) is
 \begin{equation}\label{evolution of flavor2}
{|{\nu}_{\alpha}(t)\rangle} = U_{\alpha \alpha}(t)  {|{\nu}_{\alpha}\rangle} + U_{\alpha \beta}(t) {|{\nu}_{\beta}\rangle}
 \end{equation}
where $ U_{\alpha \alpha}(t)$ and $U_{\alpha \beta}(t)$ are given by
 \begin{equation}
     U_{\alpha \alpha}(t) =\cos^2\theta \exp^{\frac{-iE_{j} t}{h}} + \sin^2{\theta}\exp^{\frac{-iE_{k}t}{h}}  , \qquad
     U_{\alpha \beta}(t) =\cos\theta \sin\theta [\exp^{\frac{-iE_{k} t}{h}} - \exp^{\frac{-iE_{j} t}{h}}]
\end{equation}
As a result, Eq.(\ref{evolution of flavor2}) can be rewritten as
\begin{equation}\label{component of flavorstate}
    {|{\nu}_{\alpha}(t)\rangle} = U_{\alpha \alpha}(t){|{\nu}_{\alpha}\rangle} +  U_{\alpha \beta}(t) {|{\nu}_{\beta}\rangle} = \begin{pmatrix}
        0\\
         U_{\alpha \alpha}(t)\\
         U_{\alpha \beta}(t)\\
         0
    \end{pmatrix}
\end{equation}
Hence, by inserting Eq.(\ref{component of flavorstate}) into Eq.(\ref{density matrix}), density matrix in Eq.(\ref{density matrix}) can be obtained as
\begin{equation}
    \rho=\begin{pmatrix}
        0 & 0 & 0 & 0\\
        0 & | U_{\alpha \alpha} |^{2} & U_{\alpha \alpha} U^{*}_{\alpha \beta} & 0\\
        0 &  U^{*}_{\alpha \alpha} U_{\alpha \beta} &  | U_{\alpha \beta} |^{2} &0\\
        0 & 0 & 0 & 0
    \end{pmatrix}
\end{equation}
where elements of the density matrix are given by
\begin{equation}\label{elements1-density matrix}
{\mid U_{\alpha \alpha} \mid}^2=\cos^4{\theta}+\sin^4{\theta}+2\cos^2{\theta}\cos{\frac{\Delta m^2 t}{2E}} 
\end{equation}  
\begin{equation}\label{elements2-density matrix}
    U_{\alpha \alpha}U_{\alpha \beta}^*=\sin {\theta}\cos{\theta}\left[\cos^2{\theta}\,e^{i\frac{\Delta m^2 t}{2E}}-\sin^2{\theta}\,e^{-i\frac{\Delta m^2 t}{2E}}+\sin^2{\theta}-\cos^2{\theta} \right] 
\end{equation}
\begin{equation}\label{elements3-density matrix}
    U_{\alpha \alpha}^*U_{\alpha \beta}=\sin {\theta}\cos{\theta} \left[\cos^2{\theta}\,e^{-i\frac{\Delta m^2 t}{2E}}-\sin^2{\theta}\,e^{i\frac{\Delta m^2 t}{2E}}+\sin^2{\theta}-\cos^2{\theta} \right] 
\end{equation}
\begin{equation}\label{elements4-density matrix}
            {\mid U_{\alpha \beta} \mid}^2=\sin^2{2\theta}\sin\frac{\Delta m^2 t}{2E}
          \end{equation}
 On the other hand, reduced density matrices of matrix $\rho$ are given by
 \begin{equation}\label{reduce density alpha}
    \rho_\alpha=\text{Tr}_\beta(\rho)={\mid U_{\alpha \alpha} \mid}^2{|{1}\rangle}{\langle {1}|}+{\mid U_{\alpha \beta} \mid}^2{|{0}\rangle}{\langle {0}|}
          \end{equation}
          
          \begin{equation}\label{reduce density beta}
    \rho_\beta=\text{Tr}_\alpha(\rho)={\mid U_{\alpha \alpha} \mid}^2{|{0}\rangle}{\langle {0}|}+{\mid U_{\alpha \beta} \mid}^2{|{1}\rangle}{\langle {1}|} .
          \end{equation}
          
The linear entropy associated with the reduced state after tracing over one mode (flavor) can be computed straightforwardly (the entropies of the muon neutrino state and the electron neutrino state yield similar results \cite{Blasone2010})

\begin{eqnarray}\label{entropy2}
S(\rho_\alpha) &=& S(\rho_\beta) \\ \nonumber
&=& -  \left(  \cos^{4}\theta + \sin^{4} \theta + 2\cos^{2} \theta \sin^{2} \theta \cos\frac{\Delta m^{2} t}{2E} \right) \\ \nonumber &\times&
\log \left( \cos^{4} \theta + \sin^{4} \theta + 2 \cos^{2} \theta \sin^{2} \theta \cos \frac{\Delta m^{2} t}{2E} \right) \\ \nonumber
&-& \left( \sin^{2} 2\theta \sin\frac{\Delta m^{2} t}{2E} \right)
\log \left(\sin^{2}2 \theta \sin\frac{\Delta m^{2} t}{2E} \right)
\end{eqnarray}
In other words, entanglement entropy is equal to:
\begin{equation}\label{entanglement-entropy}
S(\rho_\alpha)=S(\rho_\beta)= -(P_{sur})\log(P_{sur}) - (P_{osc})\log(P_{osc})
\end{equation}
where $P_{sur}$ refers to the possibility in Eq.(\ref{elements1-density matrix}) that a neutrino remains in its initial flavor state without undergoing a flavor change, and $P_{osc}$ signifies the probability in Eq.(\ref{elements4-density matrix}) that a neutrino transitions from one flavor state to another during its propagation through space. This relationship highlights how the entanglement entropy, as quantified by the Von Neumann entropy, is connected to the probabilities that describe the behavior of neutrinos in terms of flavor survival and oscillation \cite{Blasone2009}.

\section {Numerical Results}

In the previous Section, we discussed entanglement entropy between neutrino eigenstates using Neumann entropy. In this section, we explore the entanglement entropy to the phase shift of the neutrino oscillation based on Eq.(\ref{defomed-probability}). 

\begin{figure}[h]\label{normal entropy}
    \centering
    \includegraphics[width=0.6\linewidth]{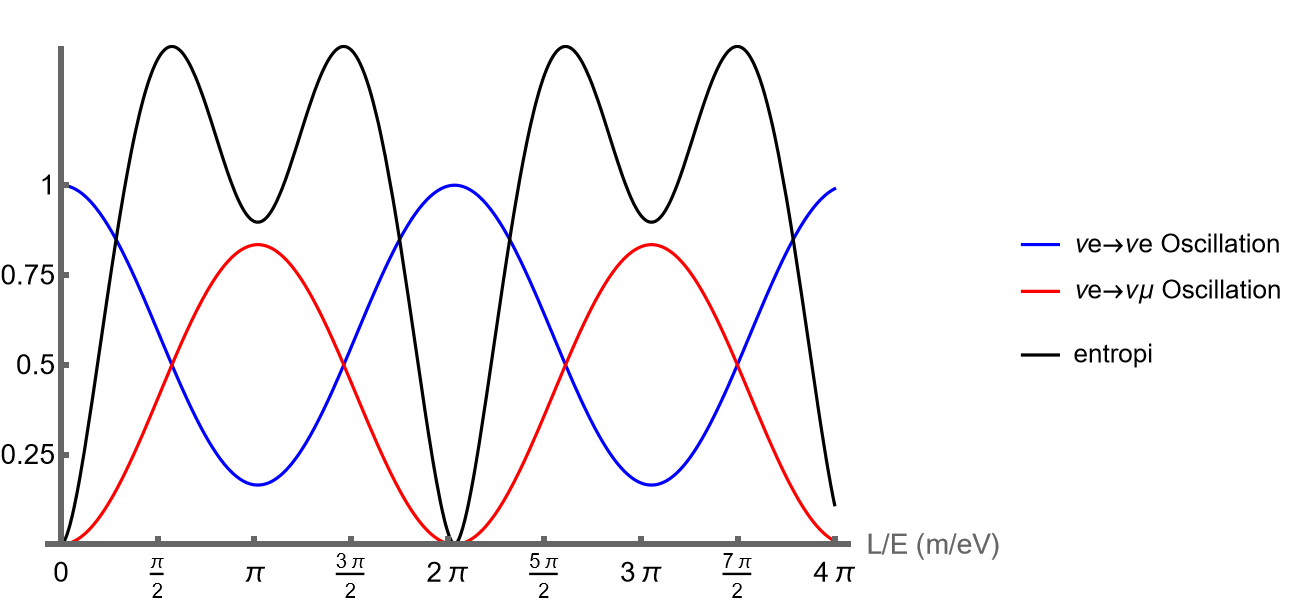}
    \caption{The blue curve denotes the survival probability $P(\nu_e\rightarrow \nu_e)$, the red curve corresponds to the transition probability from electron neutrino to a muon neutrino $P(\nu_e\rightarrow \nu_\mu)$ and the black curve corresponds to the entropy}
    \label{Figure-1}
\end{figure}

First, we consider $\nu_{e} \rightarrow \nu_{e}$ survival probability, $\nu_{e} \rightarrow \nu_{\mu}$ transition probability and entanglement entropy for the Eq.(\ref{probability}). For numerical analyses we set $\theta_{12} = 33$, $\Delta m^2_{12} =7.37\times 10^{-5}$ eV $^2$ and $c=3\times 10^8$ Km/s. The above equation expresses the fact that flavor neutrino states at any time can be regarded as entangled super-positions of the mass qubits $|{\nu_i}\rangle$, where the entanglement is a function of the mixing angle only. The entanglement entropy of the neutrino for two flavors; $\alpha = e$ and $\beta = \mu$ is plotted in Fig.\ref{Figure-1}.
In this figure, we show the behavior of $S(e,\mu)$ as a function of the scaled, dimensionless time $t$. The plots have a clear physical interpretation, at time $t = 0$, the two flavors are not mixed, the entanglement is zero, and the global state of the system is factorized. For $t \ge 0$, flavors start to oscillate, and the entanglement is maximal at the largest mixing, $P(\nu_e\rightarrow\nu_e) = P(\nu_e\rightarrow\nu_\mu)= 0.5$, and minimum at $t=\pi$. Accordingly, entropy is maximum at points where the probability of survival of an electron neutrino and the probability of oscillating a muon neutrino are equal. High values of entropy suggest greater mixing between the flavor states, indicating a more entangled quantum system. In the area between these points, that is, in the range where the probability of oscillation is greater than the probability of survival, entropy decreases again. At points where the probability of survival is maximum, entropy takes its minimum values.

\begin{figure}[h]   
        \centering
         \includegraphics[width=0.4\linewidth]{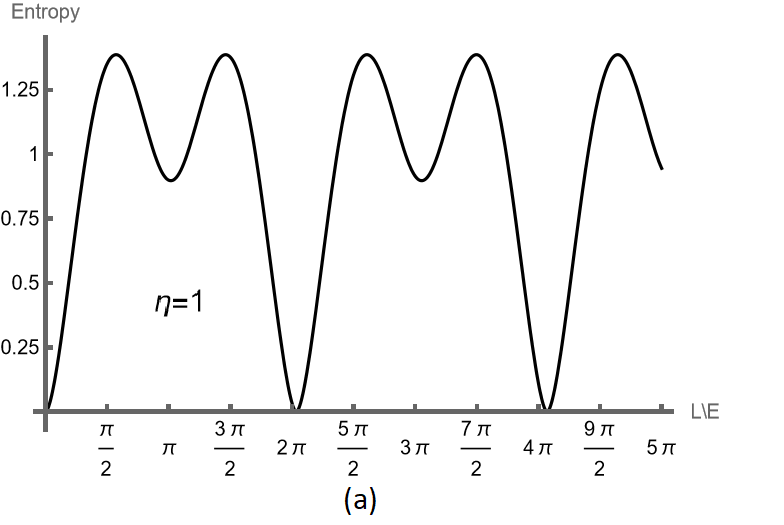}\quad
         \includegraphics[width=0.4\linewidth]{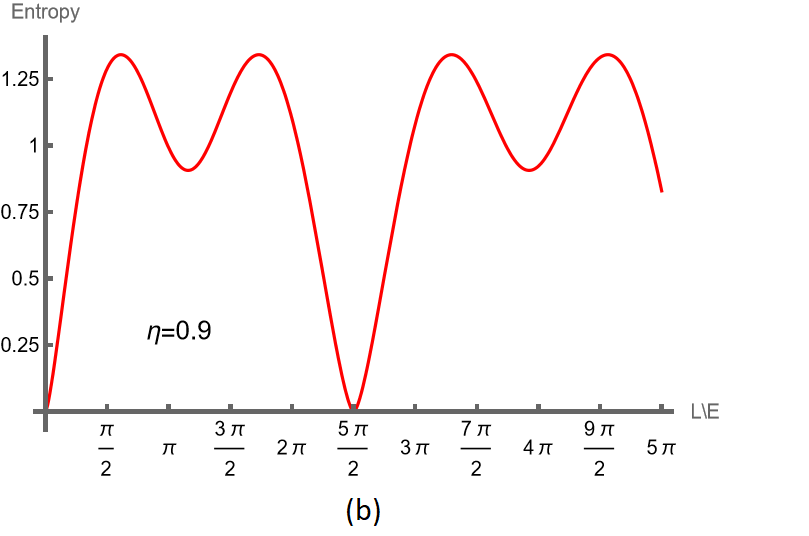}
        \medskip
         \includegraphics[width=0.4\linewidth]{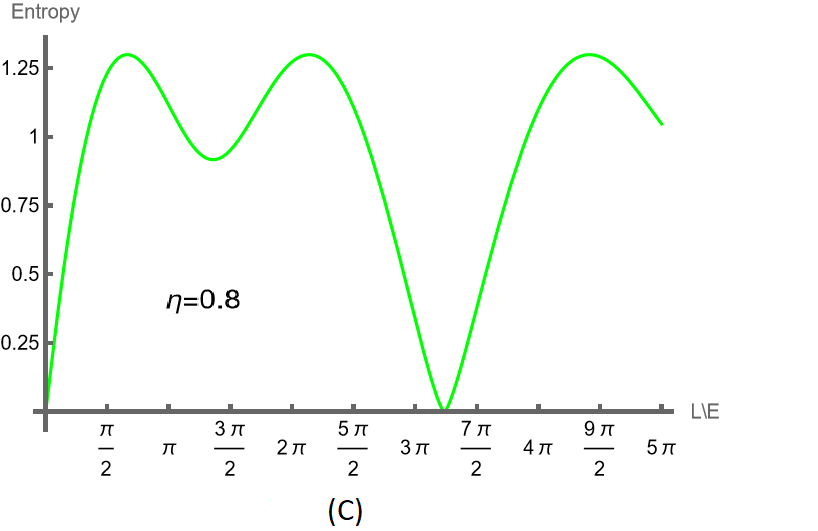}\quad
         \includegraphics[width=0.4\linewidth]{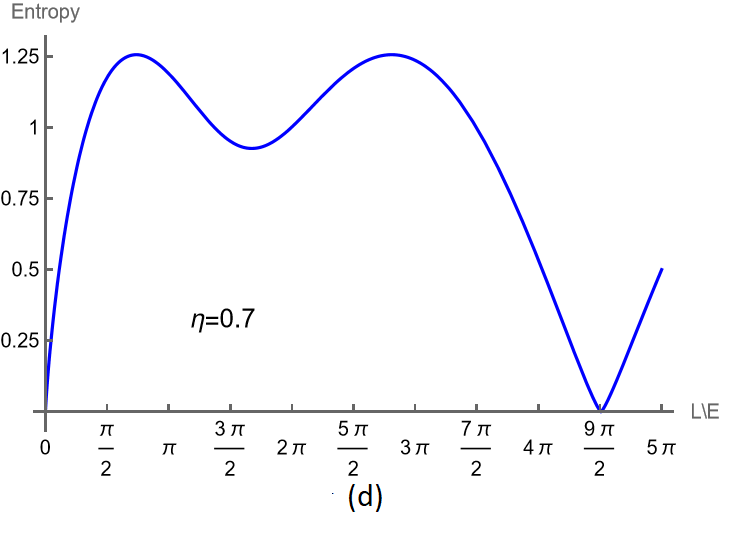}
    \caption{In (a), the entropy of neutrino oscillation $S(e,\mu)$ Minkowski space-time for $\eta=1$ indicates the ordinary Minkowski space-time. The entropy in the deformed Minkowski space-time is given in (b–d) sub-figures. The deformation parameters of the Minkowski space-time are $\eta=0.9$ in (b) $\eta=0.8$ in (c) and $\eta=0.7$ in (d)}
    \label{Figure-2}
\end{figure}

\begin{figure}[h]   
    \centering
    \includegraphics[width=0.6\linewidth]{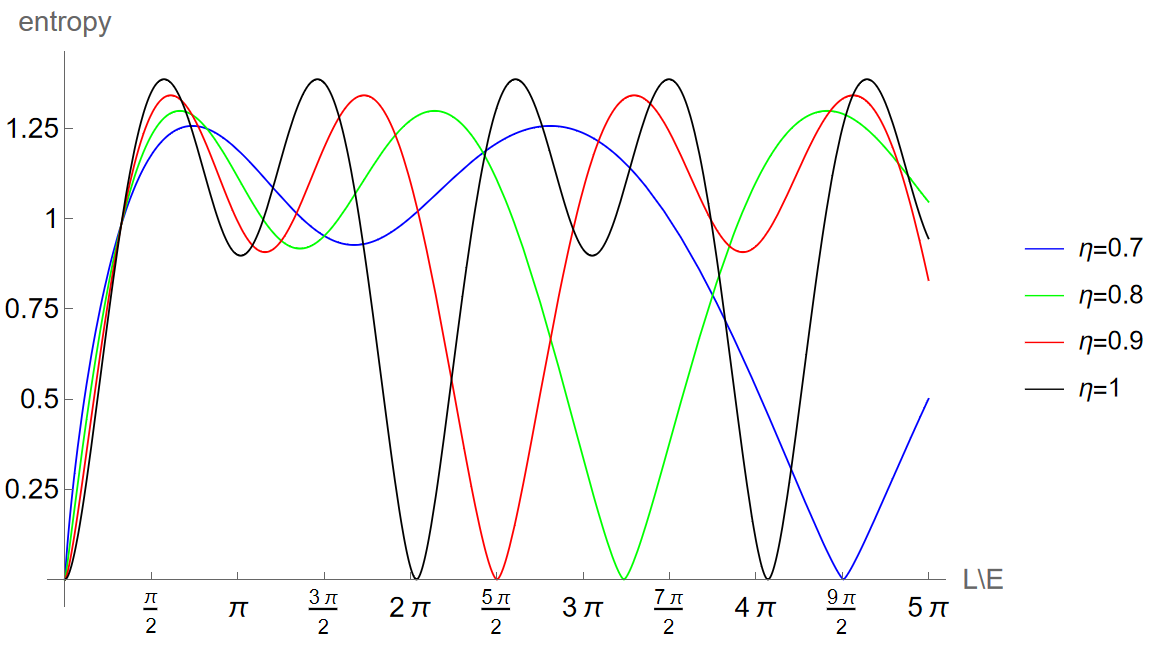}
    \caption{Comparison of the entropy of neutrino oscillation $S(e,\mu)$ for various fractional parameters}
    \label{Figure-3}
\end{figure}

\begin{figure}[h]
    \centering
    \includegraphics[width=0.7\linewidth]{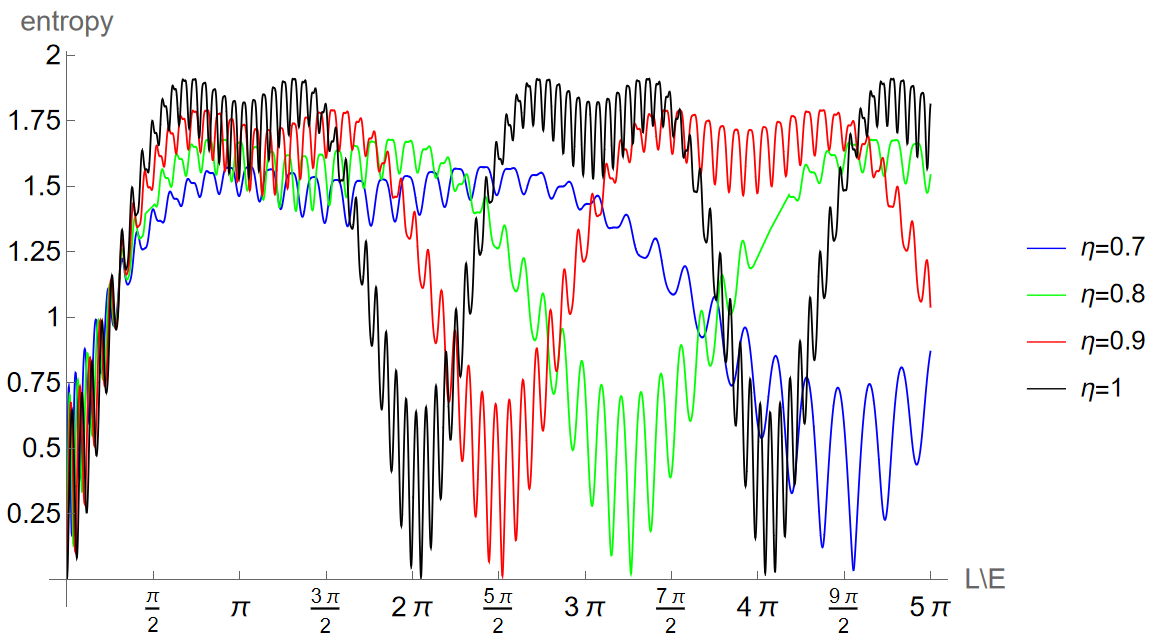}
    \caption{Comparison of the entropy of neutrino oscillation for three flavors}
    \label{Figure-4}
\end{figure}

Now, we numerically analyzed entanglement entropy by using Eq.(\ref{defomed-probability}) to see entropy change in the presence of anomalous behavior in the neutrino oscillations. By using Von Neumann entanglement entropy formalism, introduced above, we computed entanglement for various $\eta$ values. The obtained results are shown in Figs.\ref{Figure-2}, \ref{Figure-3} and \ref{Figure-4}. Firstly, we plotted the entanglement entropy results for the transition between two flavors from $\nu_{e}$ to $\nu_{\mu}$ for various $\eta$ in Figs.\ref{Figure-2} and \ref{Figure-3}. Additionally, we also presented entanglement entropy results for the transition between three flavors in Fig.\ref{Figure-4}. 

In Fig.\ref{Figure-2}(a) entanglement entropy is given for $\eta=1$. This figure is the same as the entropy given in Fig.\ref{Figure-1}. 
As seen from Fig.\ref{Figure-2}(a) the distribution of the entanglement is quite periodic and has a symmetric shape. However, for $\eta<1$ values, it can be seen that the symmetry of the entropy distribution is disrupted and the periodicity disappears. This situation becomes more evident for small values of $\eta$. For instance, for $\eta=1$ in (a) while the period of oscillation is equal to $2\pi$, for $\eta=0.9$ in (b) the period of oscillation shifts to $2.5\pi$, for $\eta=0.8$ in (c) period shifts to $3.25\pi$, and finally, for $\eta=0.7$ in (d) the period shifts to $4.5\pi$. These results indicate that the entanglement entropy of each subsequent oscillation of the neutrino increases with time in the deformed space-time depending on the deformation parameter $\eta$. Additionally, to show phase shift due to $\eta$ we plotted four entanglement figures as superimposed in Fig.\ref{Figure-3}. It was observed that the phase shift between eigenstates is not constant and gradually increases with each subsequent oscillation period. This figure may be helpful for the reader to compare the changes of entropy and phase shift, more easily.

Furthermore, we plotted entanglement entropy for three-flavor neutrino oscillations for various $\eta$ parameters as seen in Fig.\ref{Figure-4}. Here we computed entanglement entropy using three flavor transition probability. As can be seen, the results obtained for the three flavors are compatible with the results obtained for the two flavors.

\section{Conclusion}

In this Letter, we discussed entanglement entropy between eigenstates of the neutrino which moves in the deformed space-time. Before presenting the numerical results, we introduced the theoretical framework of the neutrino oscillation and the entanglement computing method for the neutrino oscillations in Section \,\ref{Section-II}. Afterward, we briefly summarized phase shift studies that may appear in the neutrino oscillation and fractional neutrino oscillation probability proposed by Aydiner in Ref.\,\cite{Aydiner}.

We numerically obtained entanglement entropy using the fractional probability function of the neutrino oscillation in Eq.(\ref{defomed-probability}) for the various values of the deformation parameter $\eta$. As seen from Fig.\ref{Figure-2}-\ref{Figure-4}, we show that entanglement entropy also shifts in the case of the presence of the phase shift in the neutrino oscillation for two and three flavor cases unlike the conventional approach given in Fig.\ref{Figure-1}. Numerical results indicate that when the space-time deformation increases, the entanglement entropy also increases and asymmetrically shifts in time like fractional transition probability in Eq.(\ref{defomed-probability}). 

The increment of the area for the subsequent oscillation in the entanglement entropy may denote the increment of the information entropy. This is the result of the phase shift in the oscillation probabilities between neutrino eigenstates. It may be early at this stage to interpret the physical counterpart of the increase in entropy with each subsequent oscillation. However, we can give a possible interpretation. Here, entanglement entropy in Eq.(\ref{entropy1}) can be considered as information entropy. As known from information theory, information entropy is equal to the thermodynamic entropy, So, the information may lead to thermodynamic works. If such a relationship is established between information and thermodynamics, a relationship can be established between the increment of the entanglement entropy and cosmology.

Finally, we have noticed that entanglement is compatible with neutrino flavor transition probabilities, as evidenced by prior studies  \cite{Blasone2008a,Blasone2009,Blasone2010,Blasone2013,Blasone2014,Blasone2014b,Blasone2015}. Similarly, in our study, we have also shown that entanglement entropy in Eq.(\ref{entanglement-entropy}) is compatible with flavor transition probability in Eq.(\ref{defomed-probability}).  


\bibliography{Neutrino}

\end{document}